\documentclass[reprint]{revtex4-1}

\usepackage{amssymb}
\usepackage{tabularx}
\usepackage{epsfig}
\usepackage{graphicx}%
\usepackage{amsmath}%
\usepackage{colortbl}

\begin{document}

\title{\boldmath Quantitative Study of Geometrical Scaling in Charm Production at HERA}

\begin{abstract}
We apply the method of ratios to search for geometrical scaling in the charm production in deep inelastic scattering. To this end we use recent combined data from H1 and ZEUS experiments. Two forms of geometrical scaling are tested: originally proposed scaling which results from Golec-Biernat--W\"{u}sthoff model and scaling motivated by a dipole representation which takes into account charm mass. It turns out that in both cases some residual scaling is present and charm mass inclusion improves scaling quality. 
\end{abstract}

\author{Tomasz Stebel}
\affiliation{M. Smoluchowski Institute of Physics, Jagiellonian University, Reymonta 4, 30-059 Krakow, Poland}
\date{\today}

\maketitle
\flushbottom

\section{Introduction}

\label{intro}

Geometrical scaling (GS) is a well-known property of Deep Inelastic Scattering (DIS) for low Bjorken $x$ variable. Introduced in Ref.~\cite{Stasto:2000er} GS is defined as dependence of virtual photon -- proton cross section $\sigma_{\gamma^{\ast}p}$ only on one dimensionless variable $\tau$ while in principle $\sigma_{\gamma^{\ast}p}$ depends on two independent variables: 
\begin{equation}
\sigma_{\gamma^{\ast}p}(x,Q^{2})=\sigma_{\gamma^{\ast}p}(\tau).
\label{GSdef}
\end{equation}

Since the discovery of GS several forms of scaling variable were postulated. In this paper we use historically first and the simplest one:
\begin{equation}
\tau_0=\frac{Q^2}{Q_0^2}\left( \frac{x}{x_{0}}\right)^{\lambda}.
\label{tau_0}%
\end{equation}
Here subscript in $\tau_0$ refers to zero quark mass (see Eq.~(\ref{tau_c_full})). In Ref.~\cite{DIS2} we showed that more sophisticated forms of scaling variable discussed in the literature are disfavored by inclusive DIS data. 

Geometrical scaling is connected to the existence of the energy scale $Q_{\text{s}}(x)$ (so-called saturation scale) which is a border between dense and dilute gluonic systems (for review see {\it e.g.} Refs.~\cite{Mueller:2001fv,McLerran:2010ub}). Since masses of light quarks are negligible, $Q_{\text{s}}$ is the only intermediate energy scale and one can argue that scaling variable, as dimensionless quantity, should be a ratio $Q^2/Q_{\text{s}}^2(x)$. In charm production GS is expected to be violated due to large quark mass $m_c \approx 1.3$ GeV. One can, however, modify ($\ref{tau_0}$) to take into account $m_c$ and obtain approximate scaling (following Ref.~\cite{Avsar:2007ht}):
\begin{equation}
\tau_c=\left(1+\frac{4 m_c^2}{Q^2} \right)^{1+\lambda} \frac{Q^2}{Q_0^2}\left( \frac{x}{x_{0}}\right)^{\lambda}.
\label{tau_c_full}%
\end{equation}

Main purpose of this article is to assess the quality of GS in the charm production. To this end we apply {\it method of ratios} developed in Ref.~\cite{DIS1} to data on $F^{c\bar{c}}_2$ (we use combined data from ZEUS and H1 experiments \cite{HERAcombinedCharm}). Analysis is performed parallelly for $\tau_0$ and $\tau_c$. Our findings can be summarized as follows: for $\tau_0$ (\ref{tau_0}) some residual scaling can be observed but its quality is low; for $\tau_c$ (\ref{tau_c_full}) scaling is clearly better but not as good as for inclusive DIS.

In Sect.~\ref{theory} we briefly present dipole description of DIS for small $x$. In Sect.~\ref{MoRdescr} we describe the method of ratios and define the criteria for GS to hold. Results for HERA data are presented in Sect.~\ref{results}. In Summary (Sect.~\ref{summary}) we compare our results with Ref.~\cite{Beuf:2008bb} where different method was applied.

\section{DIS at low x}
\label{theory}

Dipole representation provides a convenient description of DIS at small $x$. There, a virtual photon $\gamma^{\ast}$ splits into quark-antiquark pair of the transverse size $r$ which then interacts with the proton. Cross section for $\gamma^{\ast}$--proton interaction can be written as:
\begin{equation}
\sigma_{\gamma^{\ast}p}(x,Q^2)= \int \mathrm{d}^2 {\bf r} \int_{0}^{1} \mathrm{d}z | \psi({\bf r},z,Q^2)|^2 \sigma_{dp}(x,{\bf r})
\label{factFormula}
\end{equation}
where $z$ is a fraction of longitudinal photon momentum carried by quark.

First factor of the integrand in (\ref{factFormula}), so-called {\it photon wave function} $|\psi|^2$, describes photon dissociation into a dipole and can be calculated perturbatively. It is convenient to consider photon with given polarization (transverse or longitudinal): $|\psi|^2=|\psi_T|^2+|\psi_L|^2$. In the leading order: 

\begin{multline}
|\psi_{T}(\textbf{r},z,Q^2)|^2 = \frac{3 \alpha_{\rm{em}}}{2 \pi^2} \sum_f e_f^2 \: \bigg\{ \left[z^2+(1-z)^2\right] \\
\times \xi_f(z,Q^2) Q^2 K_1^2 \left( \sqrt{\xi_f(z,Q^2) r^2 Q^2} \right) \\
+ m_f^2 K_0^2 \left( \sqrt{\xi_f(z,Q^2) r^2 Q^2} \right) \bigg\}
\label{psiTform}
\end{multline}
and
\begin{multline}
|\psi_{L}(\textbf{r},z,Q^2)|^2 = \frac{3 \alpha_{\rm{em}}}{2 \pi^2} \sum_f e_f^2 \: \bigg\{ 4 Q^2 z^2(1-z)^2 \\
\times K_0^2 \left( \sqrt{\xi_f(z,Q^2) r^2 Q^2} \right) \bigg\}
\label{psiLform}
\end{multline}
where $K_{0,1}$ are the Bessel functions, sum $\sum_f$ runs over all quark flavors with charge $e_f$ and mass $m_f$; we have also introduced:
\begin{equation}
\xi_f(z,Q^2) = z(1-z)+\frac{m_f^2}{Q^2}.
\label{ksidef}
\end{equation}

Dipole cross section $\sigma_{dp}$ in (\ref{factFormula}) characterizes the interaction of quark-antiquark dipole with the proton. Golec-Biernat and W\"{u}sthoff (GBW) \cite{GolecBiernat:1998js,GolecBiernat:1999} postulated that in $\sigma_{dp}$ dipole size $r$ is scaled by saturation scale $Q_{\text{s}}$: $\sigma_{dp}(x,r)= \sigma_{dp} \left( r Q_{\text{s}}(x) \right)$ and
\begin{equation}
Q_{\text{s}}^{2}(x)=Q_{0}^{2}\left( \frac{x}{x_{0}}\right) ^{-\lambda}.
\label{Qsat}%
\end{equation}
Parameters $Q_{0}$ and $x_{0}$ set dimension and absolute value of the saturation scale. Exponent $\lambda$ governs $x$ behavior of $Q_{\text{s}}^{2}$ ($\lambda\sim 0.3$).

Changing integration variable $r\rightarrow u/Q_{\text{s}}(x)$ in Eq.~(\ref{factFormula}) one obtains:
\begin{widetext}
\begin{multline}
\sigma_{\gamma^{\ast}p}(x,Q^2)
= \frac{3 \alpha_{\rm{em}}}{\pi} \sum_f e_f^2 \int_{0}^{\infty} \mathrm{d} u \int_{0}^{1} \mathrm{d}z~ u \left\{ \left[z^2+(1-z)^2\right] \xi_f(z,Q^2) \frac{Q^2}{Q^2_{\text{s}}(x)} K_1^2 \left( \sqrt{\xi_f(z,Q^2) \frac{Q^2}{Q^2_{\text{s}}(x) }} u \right) \right. \\
\left.+ \left[ 4 z(1-z) \cdot z(1-z)+ \frac{m_f^2}{Q^2} \right] \frac{Q^2}{Q^2_{\text{s}}(x) } K_0^2 \left( \sqrt{\xi_f(z,Q^2) \frac{Q^2}{Q^2_{\text{s}}(x) }} u \right) \right\} \sigma_{dp} (u)
\label{sigmaFull}
\end{multline}
\end{widetext}

For inclusive DIS one can neglect quark masses $m_f$ since contribution to the cross section coming from light quarks (with $m_f^2 \ll Q^2$) is dominant. Then $\xi_f(z,Q^2) = z(z-1)$ and the right hand side of (\ref{sigmaFull}) depends only on one dimensionless quantity:
\begin{equation}
\tau_0=\frac{Q^2}{Q_{\text{s}}^{2}(x)}=\frac{Q^2}{Q_0^2}\left( \frac{x}{x_{0}}\right)^{\lambda}.
\end{equation}
This is exactly definition of GS with the explicit form of saturation scale (\ref{Qsat}). 

In the case of charm production, when a dipole consists of charm quarks, mass cannot be neglected any more. The cross section in this case, which we denote as $\sigma^{c\bar{c}}$, corresponds to one term from the sum in (\ref{sigmaFull}). It is clear that now left hand side of (\ref{sigmaFull}) depends not only on $\tau_0$ but also on the ratio $m_c^2/Q^2$ and GS is violated. However, since for large $Q^2$ this ratio is small one can generalize scaling variable to obtain approximate scaling. Indeed, for $z=1/2$ factor $4z(1-z)$ equals 1 and square bracket in the second term is equal $\xi_c(z,Q^2)$ (see Eq. (\ref{ksidef})). Therefore, for $z \approx 1/2$ integrand in (\ref{sigmaFull}) depends on 
$[z(1-z)+m_f^2/Q^2]Q^2/Q^2_{\text{s}}(x)$ what suggests the following scaling variable: 
\begin{equation}
\tau=\left(1+\frac{4 m_c^2}{Q^2} \right) \frac{Q^{2}}{Q_{\text{s}}^{2}(x)}
\label{tau_bez_bar}%
\end{equation}

In saturation model for low $Q^2$ one should take into account quark mass. To this end we shall follow Ref.~\cite{GolecBiernat:1998js} and replace
\begin{equation}
x\rightarrow \left(1+\frac{4 m_c^2}{Q^2} \right)x
\label{x_bar}%
\end{equation}
in the formula for saturation scale (\ref{Qsat}).

Finally, substituting (\ref{Qsat}) into (\ref{tau_bez_bar}) and raplacing Bjorken-$x$ according to (\ref{x_bar}) one obtains scaling variable for charm production:
\begin{equation}
\tau_c=\left(1+\frac{4 m_c^2}{Q^2} \right)^{1+\lambda} \frac{Q^2}{Q_0^2}\left( \frac{x}{x_{0}}\right)^{\lambda}.
\label{tau_c_full2}%
\end{equation}
Let us emphasize that in this case scaling is not exact since we fixed $z=1/2$, whereas the cross section contains integration over $\mathrm{d}z$.

\section{Method of ratios}
\label{MoRdescr}

Geometrical scaling hypothesis for charm production means that
\begin{equation}
\sigma^{c\bar{c}}(x,Q^2)=\frac{1}{Q_0^2} f(\tau) 
\label{scal_hyps}
\end{equation}
where $f$ is a universal, dimensionless function of scaling variable. Constant $Q_0^2$ sets the dimension.

Photon-proton cross section for charm production is related with structure function by simple relation: 
$\sigma^{c\bar{c}}(x,Q^2)=4\pi^2 \alpha_{em}F_2^{c\bar{c}}(x,Q^2)/Q^2$. For $F_2^{c\bar{c}}$ we shall use most recent results from HERA \cite{HERAcombinedCharm}. Measured values (in Ref.~\cite{HERAcombinedCharm} denoted as $\sigma_{\text{red}}$) are not exactly $F_2^{c\bar{c}}$ -- contribution from longitudinally polarized photons is not included and production of charm in final state is not excluded. These corrections are at most of the order of a few percent \cite{Daum,Forte} {\it i.e.} smaller than $F_2^{c\bar{c}}$ errors and should not affect our findings. 

Let us define $\sigma^{c\bar{c}}_x(Q^2):=\left. \sigma^{c\bar{c}}(x,Q^2)\right|_x~ $ {\it i.e.} cross section as a function of $Q^2$ with fixed $x$. Eq.~(\ref{scal_hyps}) means that functions $\sigma^{c\bar{c}}_x(\tau)$ for different $x$ follow one curve. In fact instead of fixing $x$ one can use any other kinematical variable $v$ and consider $\sigma_v$ evaluated in terms of $Q^2$ or $\tau$. It turns out that here the most convenient choice for $v$ is a photon-proton center of mass energy:
\begin{equation}
W=\sqrt{\frac{Q^2}{x}-(Q^2-M_p^2)}
\end{equation}
where $M_p$ is a proton mass.

We define:
\begin{equation}
\sigma^{c\bar{c}}_W(Q^2):=\left. \sigma^{c\bar{c}}(x,Q^2) \right|_{W=\sqrt{Q^2/x-(Q^2-M_p^2)}}
\label{Wspectradef}
\end{equation}
{\it i.e.} cross section as a function of $Q^2$ with fixed $W$. As previously, GS is satisfied when functions $\sigma^{c\bar{c}}_W$ evaluated in terms of $\tau$ follow one curve for all $W's$ (see Fig.~\ref{figGS}). This observation is essential in our method: one can construct ratios $\sigma^{c\bar{c}}_{W_1}(\tau)/\sigma^{c\bar{c}}_{W_2}(\tau)$ and check whether they are equal 1.

Before we proceed, let us discuss the choice of kinematical variable $v$ which we fix to construct the ratios. In Ref.~\cite{DIS1} we performed analysis of GS in inclusive DIS both with fixed $x$ (so-called Bjorken $x$ binning) and $W$ (energy binning). Although both approaches give similar results the latter is disfavored since HERA data are provided in $(x,Q^2)$ bins -- changing bins into $(W,Q^2)$ introduces additional errors. In Ref.~\cite{Stebel:2012ky} details of bins change has been presented. On the other hand points with fixed energy span over much wider $Q^2$ range what simplifies analysis significantly. Moreover, it has been shown that GS is exhibited by $p_\mathrm{T}$ spectra in high-energy hadronic collisions (see Refs.~\cite{McLerran:2010ex,Praszalowicz:2011tc,Praszconf}) where the energy binning is natural.

HERA data for charm production are much poorer than for inclusive DIS. $\sigma^{c\bar{c}}_x(Q^2) $ consist of few points and ratios cannot be constructed. This force us to use energy binning even though data are provided in $(x,Q^2)$ bins. Here we use the same binning as in Refs.~\cite{DIS1} and \cite{Stebel:2012ky}. Limiting values $W^{\prime}_{\mathrm{min}}$, $W^{\prime}_{\mathrm{max}}$ are shown in Table \ref{table1}. Energy $W$ is taken as a mean of these values. Number of points in each bin is also displayed.

\begin{table}[ptb]
\begin{center}%
\begin{tabular}[c]{|c|c|c|c|c|c|c|}\hline
$W^{\prime}_{\mathrm{min}}\mathrm{\ [GeV]}$ & 62.7 & 81.6 & 106 & 137.9 & 179.2 & 233 \\ \hline
$W^{\prime}_{\mathrm{max}}\mathrm{\ [GeV]}$ & 81.6 & 106 & 137.9 & 179.2 & 233 & 302.9 \\ \hline
$W\mathrm{\ [GeV]}$ & 72.2 & 93.8 & 122 & 158.5 & 206.1 & 267.9 \\ \hline
Number of points & 7 & 6 & 10 & 9 & 11 & 3 \\ \hline
\end{tabular}
\end{center}
\caption{Limiting values for energy bins, values of energies and number of points in bins for charm production.}%
\label{table1}%
\end{table}

To construct ratios of spectra (\ref{Wspectradef}) we choose {\it reference energy} $W_{\text{ref}}$ and divide $\sigma^{c\bar{c}}_{W_{\text{ref}}}$ by spectrum with different energy $W_i$: 
\begin{equation}
R_{W_{i}}(\lambda;k):=\frac{\sigma^{c\bar{c}}_{W_{\text{ref}}}(\tau_k)}{\sigma^{c\bar{c}}_{W_i}(\tau_k)} \; \text{with}\; \tau_k=\tau(W_{i},Q_{k,i}^2;\lambda)
\label{ratiodef}%
\end{equation}
where $\tau(W_i,Q_{k,i}^2;\lambda)$ is a value of scaling variable evaluated for $k$-th point of $\sigma^{c\bar{c}}_{W_i}$. Note that in general $\tau(W_{\text{ref}},Q_{k,\text{ref}}^2;\lambda) \neq \tau(W_{i},Q_{k,i}^2;\lambda)$ so to evaluate numerator of (\ref{ratiodef}) one needs to interpolate $\sigma^{c\bar{c}}_{W_{\text{ref}}}$ to $Q^2_{k,\text{int}}$ such that $\tau(W_{\text{ref}},Q_{k,\text{int}}^2;\lambda)=\tau_k$ (see Ref.~\cite{DIS1} for details). In what follows we use $W_{\mathrm{ref}}=206$ GeV because it gives the widest range of $\tau$ values and therefore ratios $R_{W_{i}}(\lambda;k)$ can be calculated for all $W_i$'s.

Uncertainty of ratio (\ref{ratiodef}) is given by:
\begin{multline}
\Delta R_{W_{i}}(\lambda;k)=\\
=\sqrt{
\left( \frac{\Delta \sigma^{c\bar{c}}_{W_{\text{ref}}}(\tau_k)}{\sigma^{c\bar{c}}_{W_{\text{ref}}}(\tau_k)}\right)^2 
+\left( \frac{\Delta \sigma^{c\bar{c}}_{W_i}(\tau_k)}{\sigma^{c\bar{c}}_{W_i}(\tau_k)}\right) ^2} R_{W_{i}}(\lambda;k) 
\end{multline}
where $\Delta \sigma^{c\bar{c}}=\Delta F_2^{c\bar{c}}/Q^2$ are experimental errors. Here we neglect interpolation errors and theoretical uncertainty $\delta=3\%$, introduced in Refs.~\cite{DIS1,DIS2} for inclusive DIS, as small comparing to experimental errors.

GS is expected to be present at low $x$ so it is profitable to introduce a parameter $x_{\mathrm{cut}}$ {\it i.e.} cut-off on Bjorken variable $x\leq x_{\mathrm{cut}}$.
Our aim is to find such $\lambda$ (for given energy $W_{i}$ and cut-off $x_{\mathrm{cut}}$)
that deviations $R_{W_{i}}(\lambda;\tau_{k})-1$ are minimal. For this purpose we define the chi-square function:%
\begin{equation}
\chi_{W_{i}}^{2}(x_{\mathrm{cut}};\lambda)=\frac{1}{N_{W_{i},x_{\text{cut}}%
}-1}\sum\limits_{k\in W_{i};\,x\leq x_{\mathrm{cut}}}\frac{(R_{W_{i}}%
(\lambda;\tau_{k})-1)^{2}}{\Delta R_{W_{i}}(\lambda;\tau_{k})^{2}}
\label{defchi2}%
\end{equation}
where $k\in W_{i};$ $x\leq x_{\mathrm{cut}}$ means that we sum over points
corresponding to given energy $W_{i}$, with $x$ values not larger
than $x_{\mathrm{cut}}$. $N_{W_{i},x_{\text{cut}}}$ is a number of such points. 

Function (\ref{defchi2}) is a main tool of quantitative study of GS. Let us denote the minimum of $\chi_{W_{i}}^{2}$ as $\lambda_{\mathrm{min}}\left( W_{i},x_{\mathrm{cut}} \right)$.
Uncertainty $\Delta \lambda_{\mathrm{min}}$ can be estimated by the condition:
\begin{equation}
\chi_{W_{i}}^{2}(x_{\mathrm{cut}};\lambda_{\mathrm{min}}\pm \Delta\lambda_{\mathrm{min}} )-\chi_{W_{i}}^{2}(x_{\mathrm{cut}};\lambda_{\mathrm{min}})=\frac{1}{N_{W_{i},x_{\text{cut}}}-1}.
\label{deltaLambda}
\end{equation}
It is clear that GS is satisfied when $\lambda_{\mathrm{min}}$ does not depend on energy $W$ and values of
$\chi_{W_{i}}^{2}(x_{\mathrm{cut}};\lambda_{\mathrm{min}})$ are around 1. Moreover, since exponent $\lambda$ governs behavior of saturation scale it should not depend on quark flavor. This means that the value of $\lambda$ for charm production should be similar to its value for inclusive DIS. This is yet another criterion for GS. 

To investigate GS for charm cross section we shall compare results obtained using scaling variables $\tau_0$ (\ref{tau_0}) and $\tau_c$ (\ref{tau_c_full}). Note that $\tau_0$ is obtained from $\tau_c$ by setting $m_c=0$. In what follows we shall treat $\tau_c$ and $\tau_0$ as the same scaling variable with masses 1.3 GeV and 0 respectively.

Finally, we set for simplicity $Q_0=1$ GeV and $x_0=1$. This can be done since the method of ratios is sensitive only to the relative values of scaling variable and multiplication by constant factor does not change ratios (\ref{ratiodef}). Note that $Q_0$ and $x_0$ determine value of the saturation scale (\ref{Qsat}) and can be found only using some model for $\sigma_{dp}$.

\begin{figure*}
\centering
\includegraphics[width=8cm,angle=0]{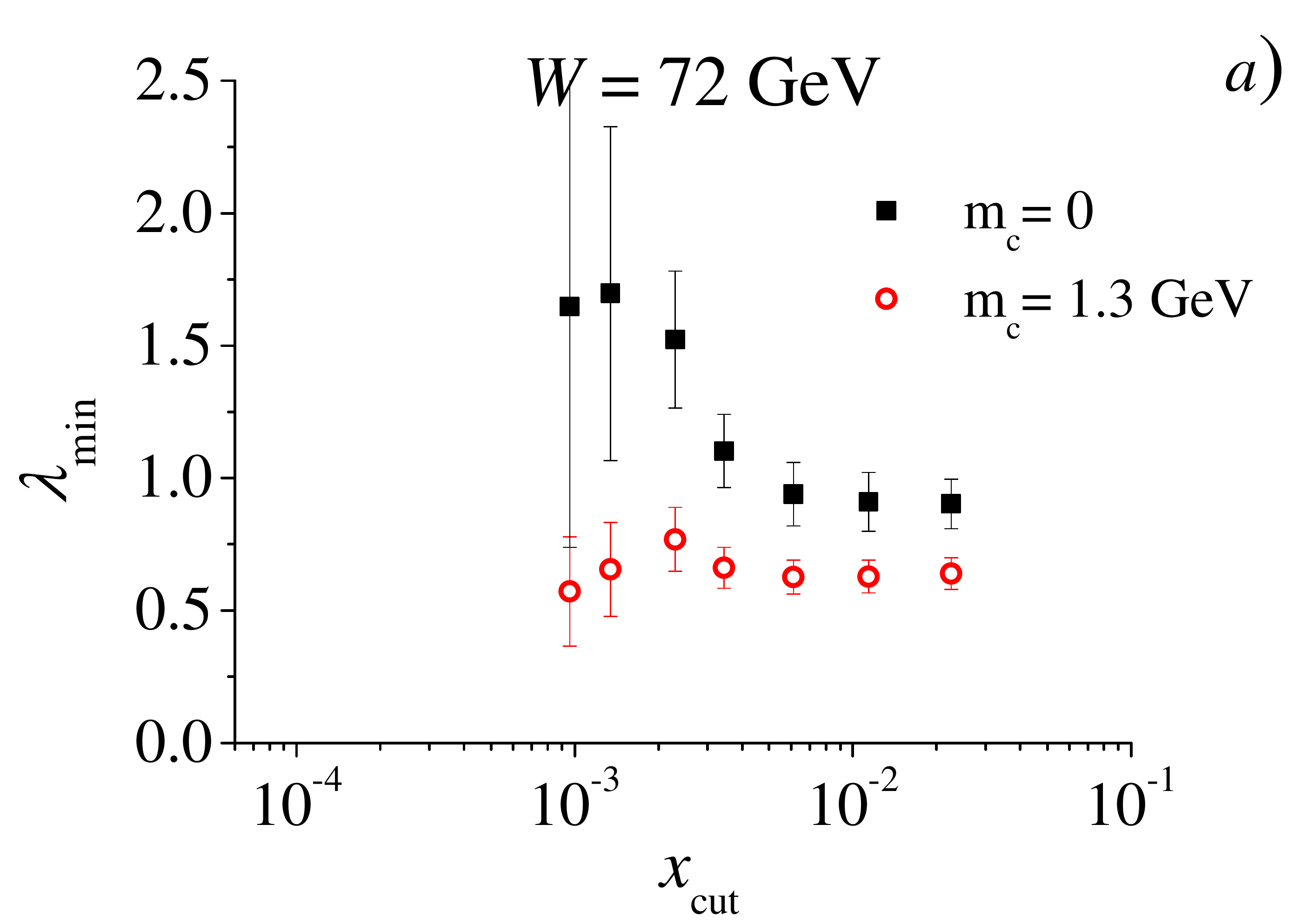}
\includegraphics[width=8cm,angle=0]{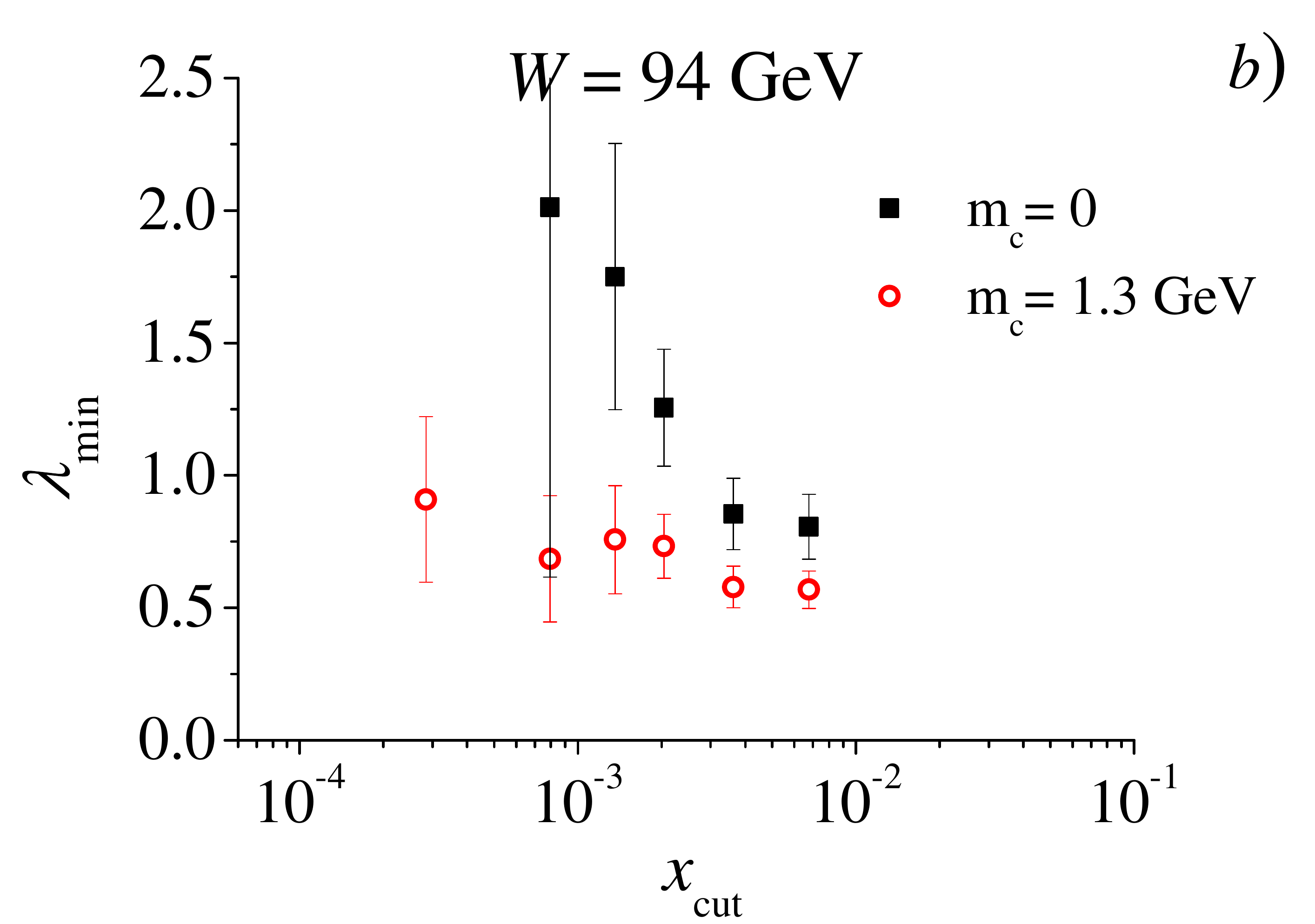}
\includegraphics[width=8cm,angle=0]{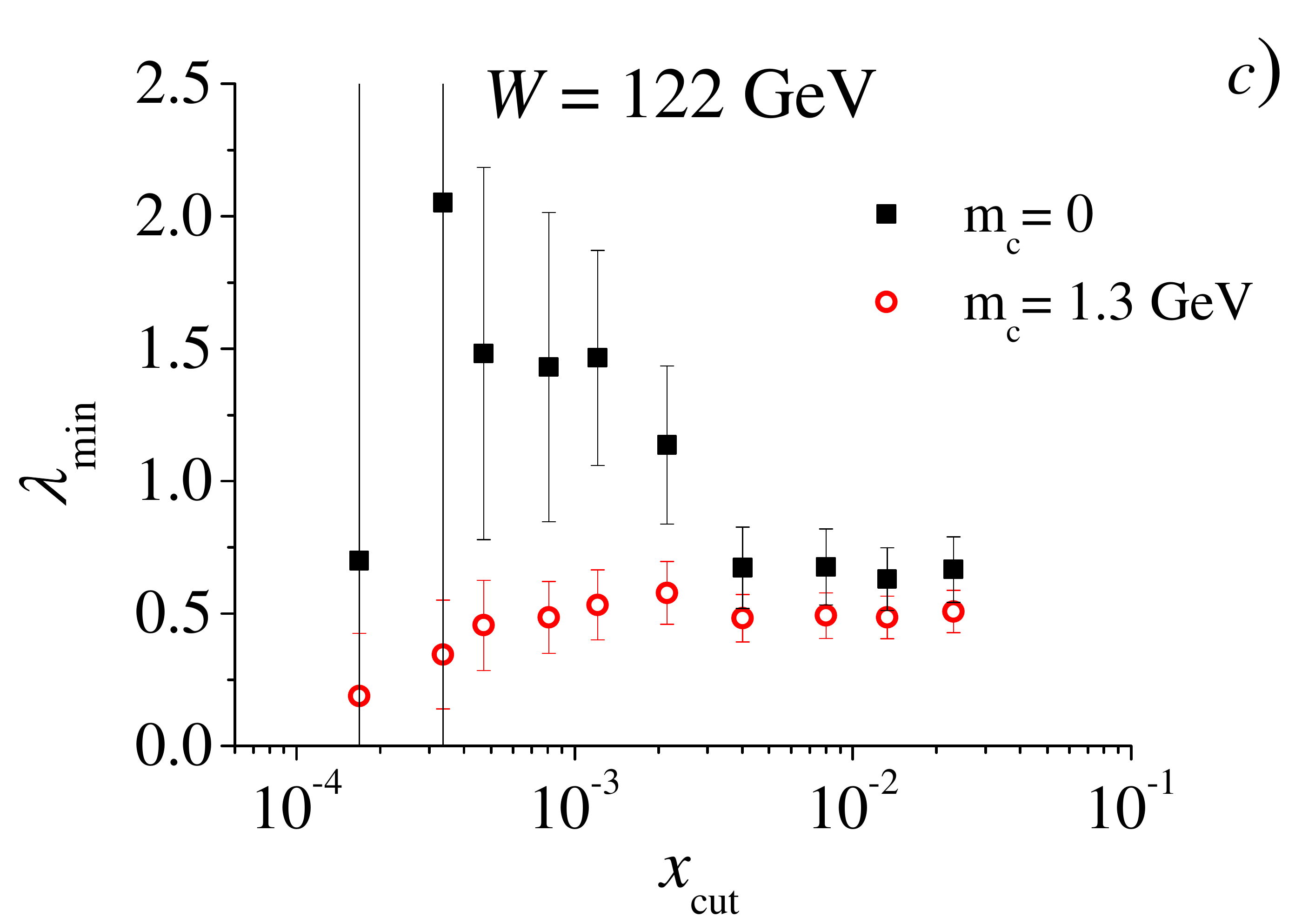}
\includegraphics[width=8cm,angle=0]{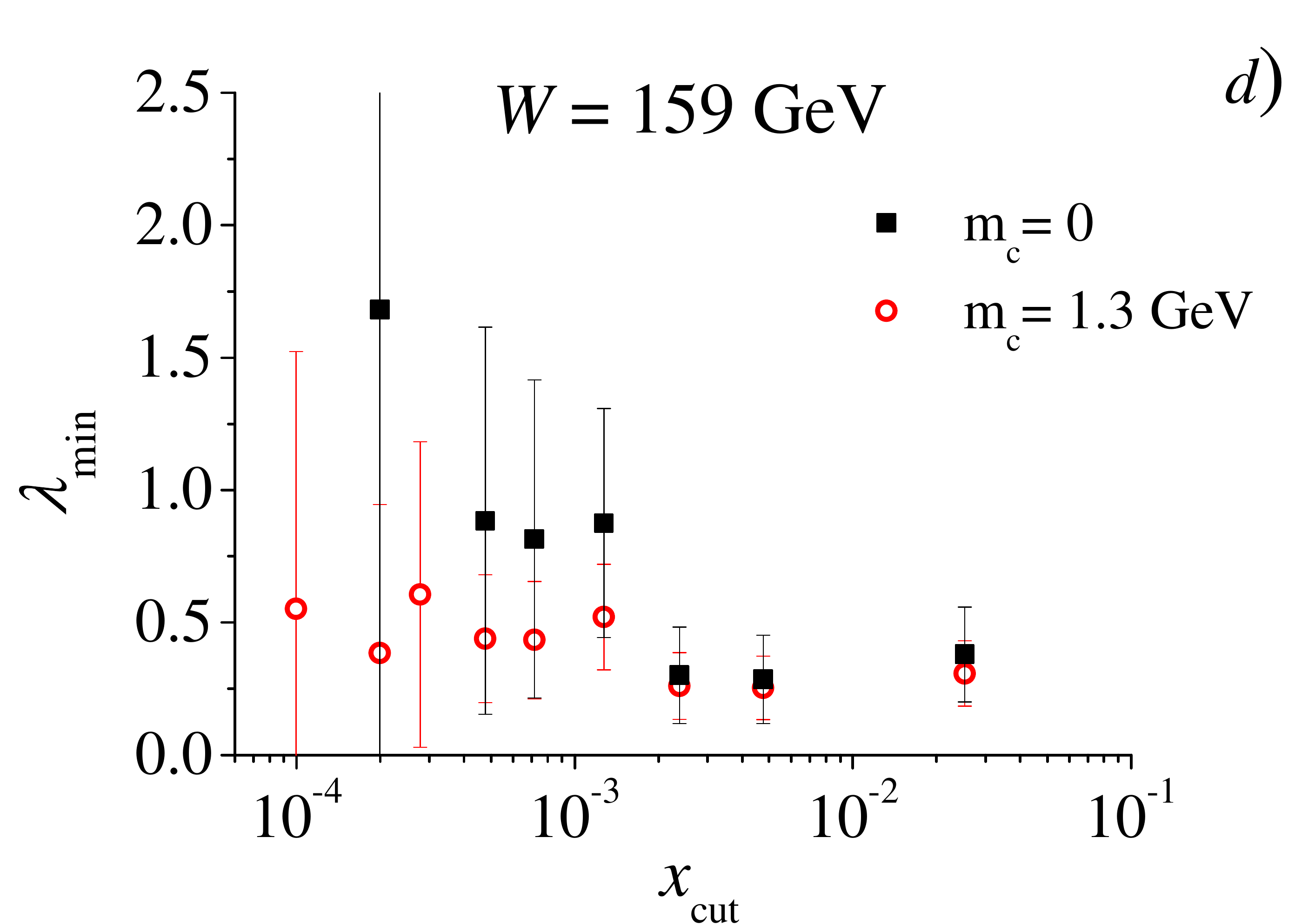}
\caption{Exponents $\lambda_{\mathrm{min}}$ as functions of $x_{\mathrm{cut}}$ for energies $W< W_{\mathrm{ref}}=206$ GeV: obtained with $m_c=0$ (black full squares) and $m_c=1.3$ GeV (red open circles).}
\label{zesWyklammin}
\end{figure*}

\begin{figure*}
\centering
\includegraphics[width=8cm,angle=0]{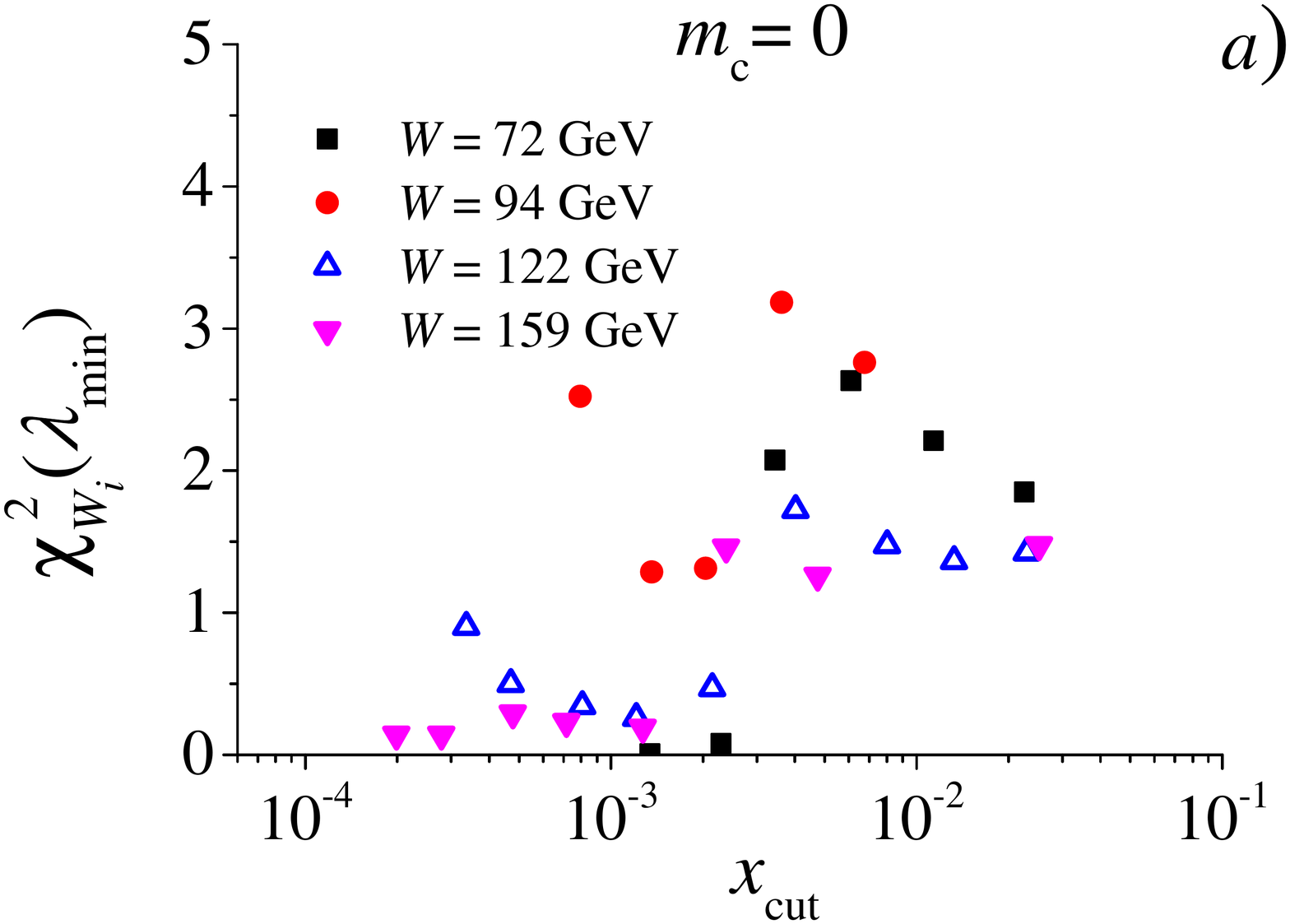}
\includegraphics[width=8cm,angle=0]{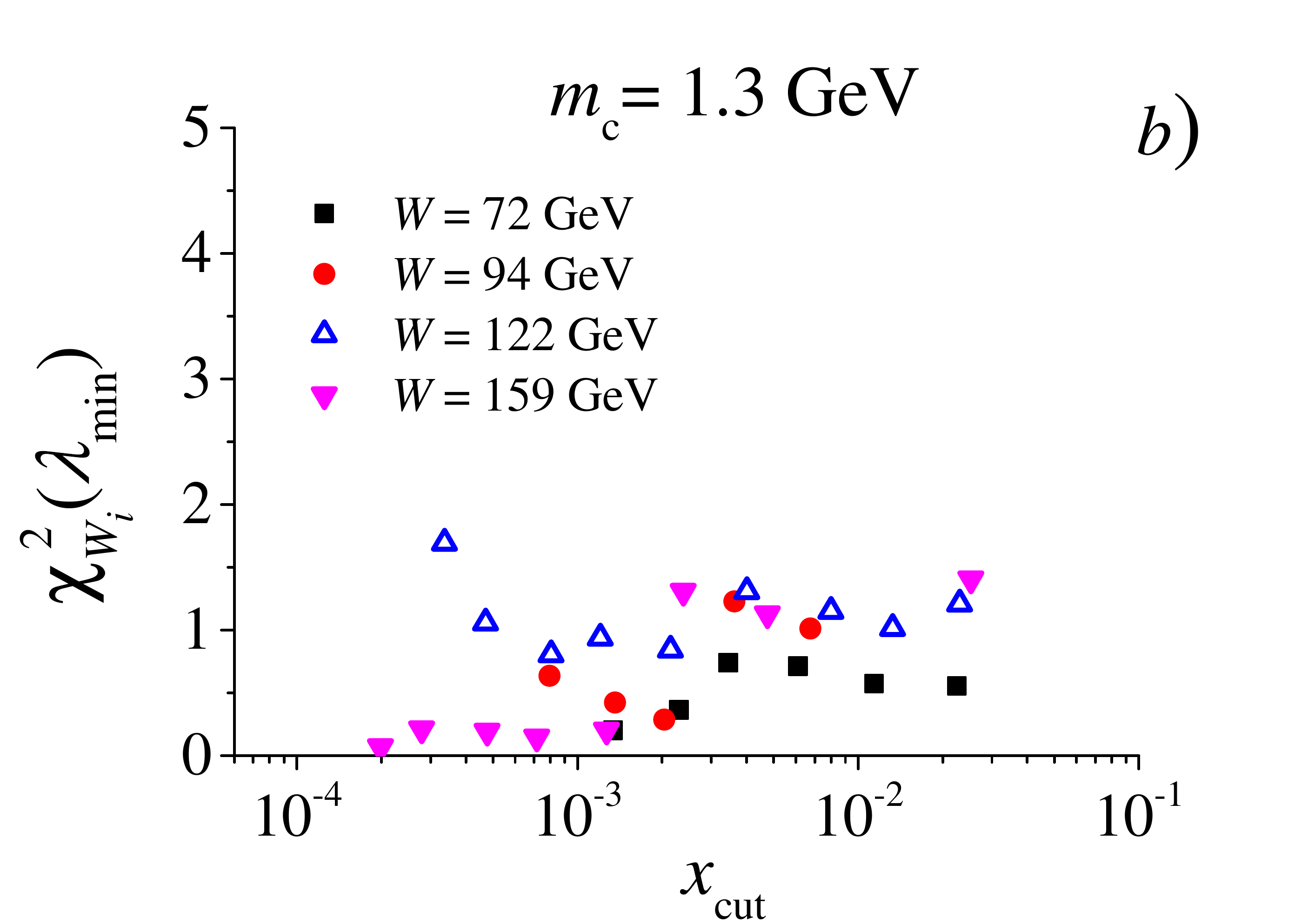}
\caption{Plots of $\chi_{W_{i}}^{2}$ for different energies as functions of $x_{\text{cut}}$: obtained with $m_c=0$ (left panel) and $m_c=1.3$ GeV (right panel).}
\label{Wchixcut}
\end{figure*}

\section{Results}
\label{results}

In Fig.~\ref{zesWyklammin} we show $\lambda_{\mathrm{min}}\left(x_{\mathrm{cut}} \right)$ for four energy bins (we omitt $W=268$ GeV since it has only 3 points and very large uncertainties). One can see that $\lambda_{\mathrm{min}}$'s obtained for $m_c=1.3$ GeV are smaller and more flat than those for $m_c=0$. Values of $\chi_{W_{i}}^{2}$ evaluated for these $\lambda_{\mathrm{min}}$'s are shown in Fig.\ref{Wchixcut}. For zero charm mass $\chi_{W_{i}}^{2}$'s are greater than 1.5 and reach 3 while for $m_c=1.3$ GeV do not exceed 1.5. This clearly shows that quality of GS is better if one include quark mass in definition of scaling variable. A word of warning is here in order. Even though $\chi_{W_{i}}^{2}$'s for $m_c=1.3$ GeV are quite small ({\it i.e.} $<1.5$) the statistical reliability of these values is low since number of points in $\chi_{W_{i}}^{2}$'s is 10 at most.

In order to check whether $\lambda_{\mathrm{min}}$ is the same for all energies we define another
chi-square function%
\begin{equation}
\tilde{\chi}^{2}(x_{\mathrm{cut}};\lambda)=\frac{1}{3}%
\sum\limits_{W_i}\frac{(\lambda_{\mathrm{min}}\left( W_{i},x_{\mathrm{cut}%
}\right) -\lambda)^{2}}{\Delta\lambda_{\mathrm{min}}\left( W_{i}%
,x_{\mathrm{cut}}\right) ^{2}}
\label{Wchitilde}%
\end{equation}
where the sum goes over four energies $W_{i}=72,94,122,159$ GeV. Minimum of (\ref{Wchitilde}) with respect to $\lambda$ is an averaged value of $\lambda_{\mathrm{min}}$ over energies which we denote by $\lambda_{\text{ave}}(x_{\text{cut}})$. The error is calculated similarly like in (\ref{deltaLambda}) {\it i.e.} by demanding that $3 \cdot \tilde{\chi}^{2}(x_{\mathrm{cut}};\lambda)$ changes by $1$ when $\lambda$ is varied around $\lambda_{\text{ave}}$. 

\begin{figure}
\centering
\includegraphics[width=8.6cm,angle=0]{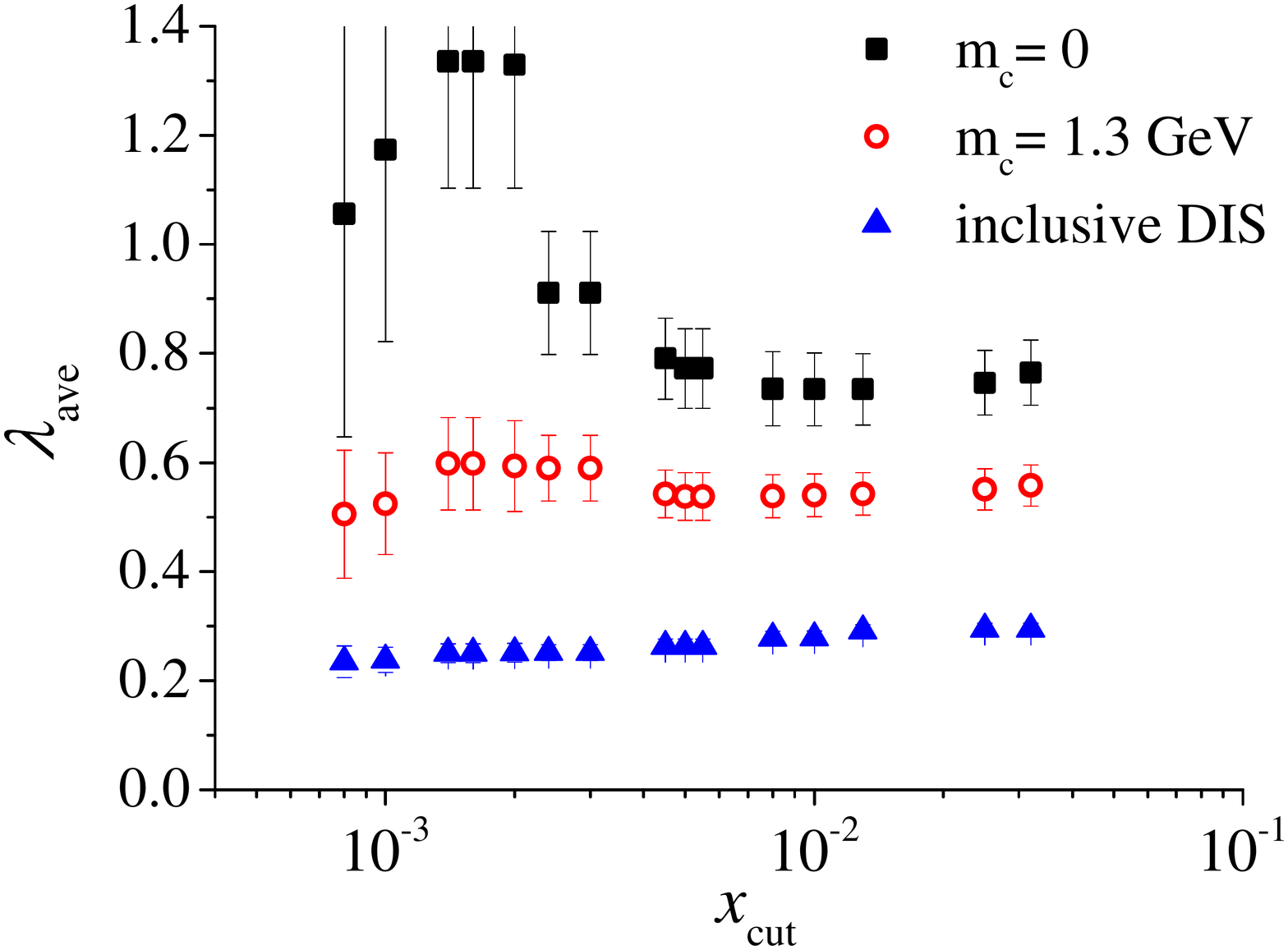}
\caption{Exponents $\lambda_{\mathrm{ave}}$ as functions of $x_{\mathrm{cut}}$ obtained for charm production data with $m_c=0$ (black squares), $m_c=1.3$ GeV (red open circles) and for inclusive DIS data (blue triangles).}
\label{lamave}
\end{figure}

In Fig.~\ref{lamave} we show comparison of $\lambda_{\text{ave}}(x_{\text{cut}})'s$ obtained using $m_c=0$ and $m_c=1.3$ GeV. We added also results for inclusive DIS data \cite{HERAcombined} restricted to the same values of $x$, $Q^2$ as used in charm production analysis. For inclusive DIS only scaling variable $\tau_0$ (\ref{tau_0}) was used. 

As optimal values of parameters $\lambda$ we take $\lambda_{\text{ave}}(x_{\text{cut}}=0.032)$. Charm production values for $m_c=0$ and $m_c=1.3$ GeV are respectively:
\begin{subequations}
\begin{eqnarray}
\lambda_0&=&0.765\pm0.06, \\
\lambda_c&=&0.558\pm0.038.
\end{eqnarray}
\end{subequations}
Inclusive DIS value:
\begin{equation}
\lambda_{\text{inc}}=0.298\pm0.011
\end{equation}
Note that errors are purely statistical. For the discussion of systematic uncertainties see Ref.~\cite{DIS1}.

One can see that exponents obtained for charm production are much larger than for inclusive DIS. This is a clear sign of GS violation. Notice that $\lambda_c$ is closer to $\lambda_{\text{inc}}$ than $\lambda_0$. This confirms our conclusion that taking into account quark mass improves GS. 

At the end let us comment on values of $\tilde{\chi}^{2}$. For $m_c=0$ we obtained $\tilde{\chi}^{2}(0.032;\lambda_0)= 2.5$, for nonzero mass $\tilde{\chi}^{2}(0.032;\lambda_c)=2.1$ and $\tilde{\chi}^{2}(0.032;\lambda_{\text{inc}})=0.7$ for inclusive DIS. This shows that GS for inclusive DIS is significantly better than for charm production as expected.

\section{Summary}
\label{summary}

\begin{figure}
\centering
\includegraphics[width=8.6cm]{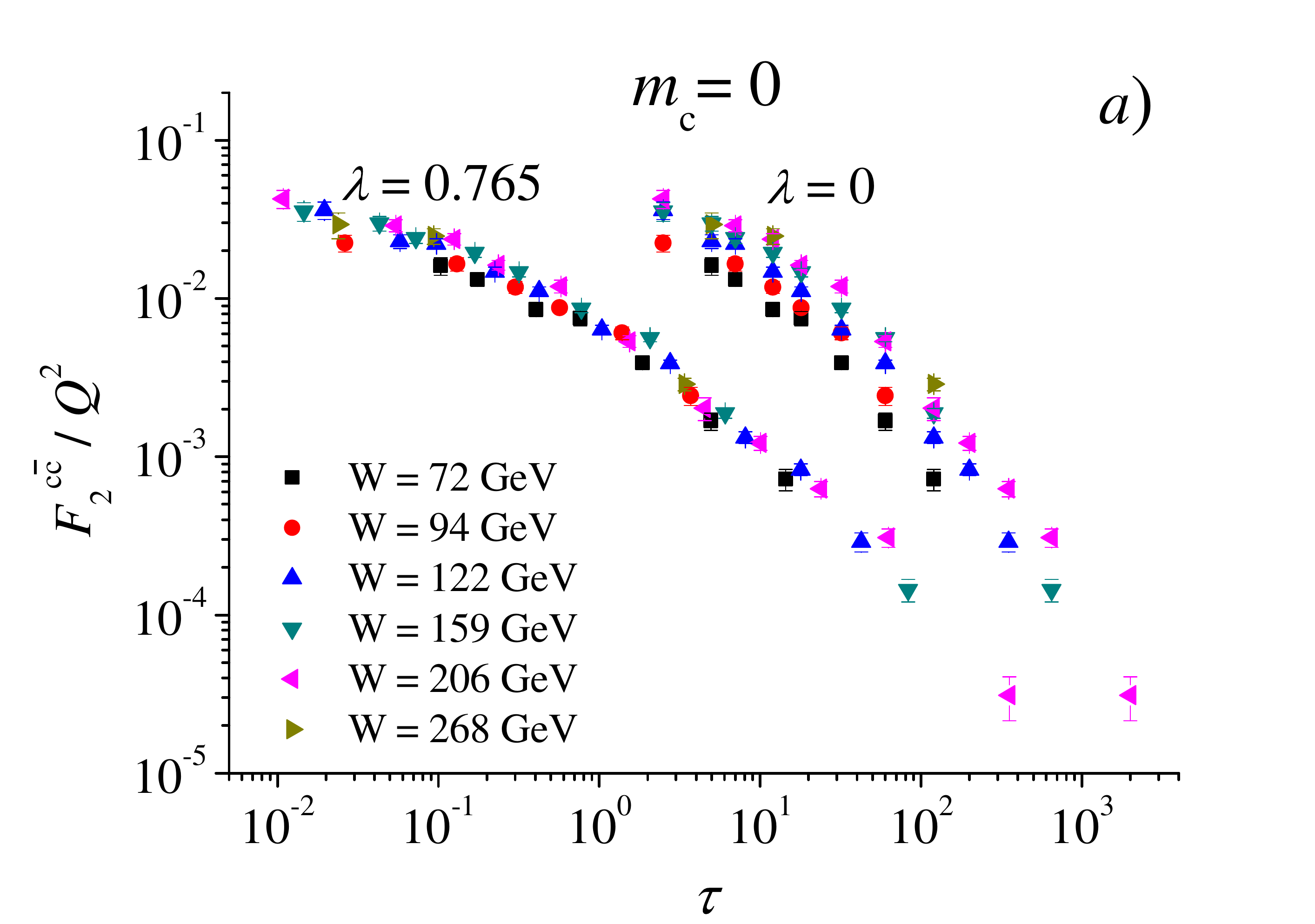}
\includegraphics[width=8.6cm]{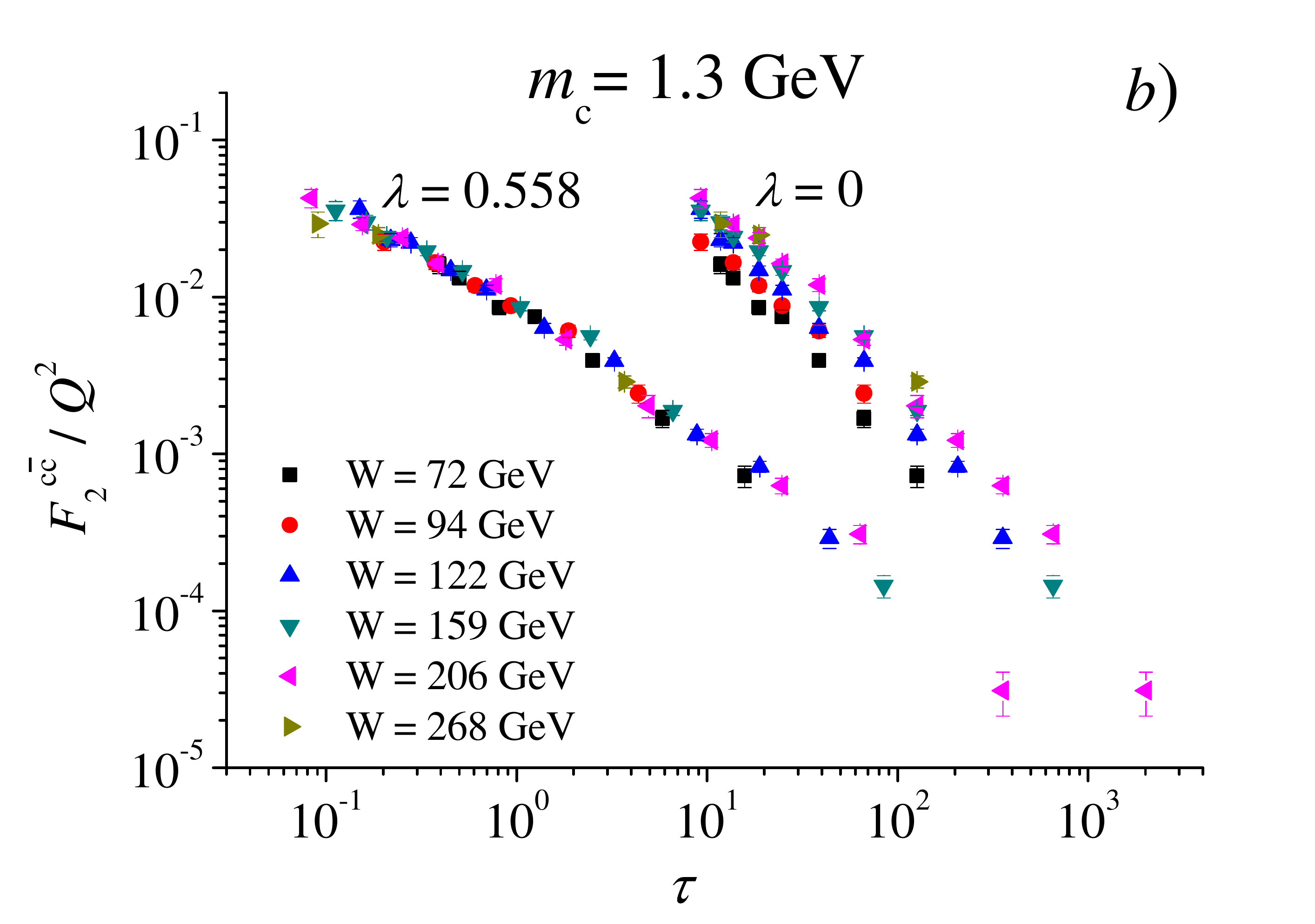}
\caption{a) $F^{c\bar{c}}_{2}/Q^{2}$ (which is $\sigma^{c\bar{c}}$ up to a constant factor) for fixed $W$'s as a function of scaling variable $\tau_c$ (\ref{tau_c_full}) with $m_c=0$. Two bands of points correspond to parameter $\lambda$ equal 0 or 0.765 (which was found in Sect.~\ref{results} as "optimal" value);\\
b) the same but for $m_c=1.3$ GeV.} 
\label{figGS}%
\end{figure}

In this paper we have considered existence of geometrical scaling for charm production in deep inelastic scattering. We have used most recent data from HERA experiment \cite{HERAcombinedCharm}. In a sense this analysis is a continuation of work presented in Ref.~\cite{DIS1} where we applied the same method to the inclusive DIS data from HERA. There, however, data are much richer and analysis was performed using also binning in Bjorken-$x$ variable. In the case of charm production we were forced to change original $x$ bins into energy bins what introduces additional errors.

Dipole approach to low $x$ DIS does not predict exact geometrical scaling in the presence of quark mass. However, since square of charm mass ($m_c^2\approx1.7$ GeV$^2$) is small comparing to the most of measured $Q^2$ values one should expect that some residual scaling can be present. Indeed, even using standard form scaling variable (\ref{tau_0}) scaling can be seen in data (upper panel in Fig.~\ref{figGS}). After taking into account charm mass (\ref{tau_c_full}) scaling is improved (lower panel in Fig.~\ref{figGS}) as one should expect knowing formula for cross section (\ref{sigmaFull}). Value of exponent: $\lambda_c=0.558$ is much larger than one obtained for inclusive DIS ($\lambda_{\text{inc}}= 0.298$). This difference is not theoretically motivated and should be treated as a consequence of scaling violation. 

Geometrical scaling in charm production (and also inclusive DIS) was tested in Ref.~\cite{Beuf:2008bb} with the help of so-called Quality Factor method. Results obtained there suggest that GS in charm production is basically the same as for the inclusive DIS. In particular for standard form of scaling variable (\ref{tau_0}) value of exponent $\lambda$ is the same for both cases $\lambda \sim 0.33$ (scaling variable (\ref{tau_c_full}) was not considered). These conclusions are far from those we have drown here. One should note, however, several important differences between both studies: in Ref.~\cite{Beuf:2008bb} authors used earlier data from HERA (not combined) and applied additional kinematical cuts to them (removed points with small $Q^2$). Moreover, values of Quality Factor as measure of scaling quality are difficult to interpret.

In Ref.~\cite{DIS1} we have established that geometrical scaling holds up to $x \sim 0.1$. This is well above domain of saturation physics ($x<0.01$) and can be predicted both by the DGLAP and BFKL evolution shames (see \cite{QCDevKwSt,QCDevIaItMc,Caola}). Here, for charm production, we have to our disposal only data with $x < 0.05$ and no sign of strong scaling violation is seen. One should expect that, similarly as in inclusive DIS, $x$ region of strong scaling violation is around $0.1$.

\section*{Acknowledgements}

Author would like to thank Michal Praszalowicz for help with preparation of this paper. This work was supported by the Polish NCN grant 2011/01/B/ST2/00492.


\begin{thebibliography}{99} 
%
\bibitem {Stasto:2000er}A.M.~Stasto, K.J.~Golec-Biernat, J.~Kwiecinski,
\emph{Geometric scaling for the total $\gamma^{*}p$ cross-section in the low x
region}, \emph{Phys.\ Rev.\ Lett.}\ \textbf{86} (2001) 596.

\bibitem{DIS2}M.~Praszalowicz, T.~Stebel, 
\emph{Quantitative Study of Different Forms of Geometrical Scaling in Deep Inelastic Scattering at HERA},
arXiv:1302.4227 [hep-ph].

\bibitem {Mueller:2001fv}A.~H.~Mueller, \emph{Parton Saturation: An Overview},
arXiv:hep-ph/0111244.

\bibitem {McLerran:2010ub}L.~McLerran, \emph{Strongly Interacting Matter
Matter at Very High Energy Density: 3 Lectures in Zakopane}, \emph{Acta
Phys.\ Pol.\ B} \textbf{41} (2010) 2799 [arXiv:1011.3203 [hep-ph]].

\bibitem{Avsar:2007ht}E.~Avsar and G.~Gustafson,
\emph{Geometric scaling and QCD dynamics in DIS},
\emph{JHEP} \textbf{0704} (2007) 067 [arXiv:hep-ph/0702087].

\bibitem{DIS1}M.~Praszalowicz and T.~Stebel,
\emph{Quantitative Study of Geometrical Scaling in Deep Inelastic Scattering at HERA},
\emph{JHEP} \textbf{03} (2013) 090 [arXiv:1211.5305].

\bibitem{HERAcombinedCharm}H.~Abramowicz \textit{et al.} [H1 and ZEUS Collaboration],
\emph{Combination and QCD analysis of charm production cross section measurements in deep-inelastic ep scattering at HERA},
\emph{Eur.\ Phys.\ J.\ C} (2013) 73:2311 [arXiv:1211.1182].

\bibitem {Beuf:2008bb}G.~Beuf, C.~Royon and D.~Salek,
\emph{Geometric Scaling of $F_2$ and $F_2^c$ in data and QCD Parametrisations},
arXiv:0810.5082 [hep-ph].

\bibitem {GolecBiernat:1998js}K.J.~Golec-Biernat, M.~Wusthoff,
\emph{Saturation effects in deep inelastic scattering at low $Q^2$ and its implications on diffraction},
\emph{Phys.\ Rev.\ D} \textbf{59} (1998) 014017 [arXiv:hep-ph/9807513].

\bibitem {GolecBiernat:1999}K.J.~Golec-Biernat, M.~Wusthoff,
\emph{Saturation in diffractive deep inelastic scattering},
\emph{Phys.\ Rev.\ D }\textbf{60} (1999) 114023 [arXiv:hep-ph/9903358].

\bibitem{Daum} K. Daum {\it et al.},
\emph{Proceedings of the workshop on ”Future physics at HERA”, eds. G.
Ingelmann, A. De Roeck and R. Klanner}, DESY, Hamburg (1996) 89 [hep-ph/9609478].

\bibitem{Forte}S.~Forte {\it et al.},
\emph{Heavy quarks in deep-inelastic scattering},
\emph{Nucl. Phys. B} \textbf{834}, (2010) 116 [arXiv:1001.2312].

\bibitem {Stebel:2012ky}T.~Stebel, Master Thesis,
\emph{Quantitative analysis of Geometrical Scaling in Deep Inelastic Scattering},
arXiv:1210.1567 [hep-ph].

\bibitem {McLerran:2010ex}L.~McLerran, M.~Praszalowicz,
\emph{Saturation and Scaling of Multiplicity, Mean $p_{\rm T}$,
$p_{\rm T}$ Distributions from 200 GeV $< \sqrt{s}$ 7 TeV},
\emph{Acta Phys.\ Pol.\ B} \textbf{41} (2010) 1917 [arXiv:1006.4293 [hep-ph]];\\
\emph{Saturation and Scaling of Multiplicity, Mean $p_{\rm T}$,
$p_{\rm T}$ Distributions from 200 GeV $< \sqrt{s}$ 7 TeV -- Addendum},
\emph{Acta Phys.\ Polon.\ B} \textbf{42} (2011) 99 [arXiv:1011.3403
[hep-ph]].

\bibitem {Praszalowicz:2011tc}M.~Praszalowicz,
\emph{Improved Geometrical Scaling at the LHC},
\emph{Phys.\ Rev.\ Lett.}\ \textbf{106} (2011) 142002 [arXiv:1101.0585 [hep-ph]].

\bibitem {Praszconf}M.~Praszalowicz,
\emph{Geometrical Scaling in Hadronic Collisions}, \emph{Acta Phys.\ Polon.\ B} \textbf{42} (2011) 1557 [arXiv:1104.1777 [hep-ph]];\\
\emph{New Look at Geometrical Scaling}, [arXiv:1112.0997 [hep-ph]];\\
\emph{Geometrical Scaling in High Energy Hadronic Collisions},
arXiv:1205.4538 [hep-ph].

\bibitem {HERAcombined}F.~D.~Aaron \textit{et al.} [H1 and ZEUS
Collaboration],
\emph{Combined Measurement and QCD Analysis of the Inclusive ep Scattering Cross Sections at HERA},
\emph{JHEP} \textbf{1001} (2010) 109, [arXiv:0911.0884 [hep-ex]].

\bibitem{QCDevKwSt}J.~Kwiecinski and A.~M.~Stasto
\emph{Geometric scaling and QCD evolution}, \emph{Phys. Rev. D} \textbf{66},
014013 (2002) and
\emph{Large geometric scaling and QCD evolution}, \emph{Acta Phys. Polon. B} \textbf{33},
3439 (2002).

\bibitem{QCDevIaItMc}E.~Iancu, K.~Itakura and L.~McLerran, 
\emph{Geometric scaling above the saturation scale}, \emph{Nucl. Phys. A} \textbf{708}, 327 (2002).

\bibitem{Caola} F.~Caola and S.~Forte, 
\emph{Geometric Scaling from DGLAP evolution}, \emph{Phys. Rev. Lett.} \textbf{101}, 022001 (2008).

\end{thebibliography}
\end{document}